\documentclass[slac_one]{revtex4}

\usepackage{graphicx}
\usepackage{fancyhdr}
\begin{document}

%Title of paper
\title{High-statistics measurement of neutral-pion pair production in 
two-photon collisions}  %% Paper title goes here

% Repeat the \author .. \affiliation  etc. as needed
%
% \affiliation command applies to all authors since the last
% \affiliation command. The \affiliation command should follow the
% other information

\author{S.~Uehara$^1$ and Y.~Watanabe$^2$ (for the Belle Collaboration)}
\affiliation{$^1$ KEK, High Energy Accelerator Research Organization, Tsukuba, 
Japan.\\
%IPNS-KEK, 
%Institute of Particle and Nuclear Studies,\\
%\hspace*{2mm} 
$^2$ Faculty of Engineering, Kanagawa University, Yokohama, Japan.}

\begin{abstract}
We have measured the cross section for $\pi^0\pi^0$ production in
  two-photon collisions using a data sample with an
  integrated luminosity of 95~fb$^{-1}$ collected with the Belle detector.
 We find at least four resonant structures including a peak from $f_0(980)$, 
and a fitting of the result to partial-wave amplitudes assuming some 
resonances are
applied.  We  also make a  discussion on the higher-energy region data
based on the preliminary results from an increased sample corresponding to 
the integrated luminosity of 223~fb$^{-1}$.
\end{abstract}

%\maketitle must follow title, authors, abstract
\maketitle

\thispagestyle{fancy}

% body of paper here - Use proper section commands
% References should be done using the \cite, \ref, and \label commands
% Put \label in argument of \section for cross-referencing
%\section{\label{}}

\section{Introduction}
  Measurements of exclusive hadronic final states in two-photon
collisions provide valuable information about the 
physics of light and heavy-quark
resonance production, perturbative and non-perturbative QCD
and hadron-production mechanism.
For the low energy region (the two-photon center-of-mass energy range, 
$W < 0.8$~GeV), 
it is expected that different electric charges of the mesons
play an essential role in the difference of the cross sections
between the $\pi^+\pi^-$ and $\pi^0\pi^0$ processes. 
Predictions are not straightforward because of non-perturbative effects.
For the intermediate energy region, formation
of meson resonances decaying to $\pi\pi$ is the dominant
contribution.  We can restrict  $I^GJ^{PC}$ 
of the $q\bar{q}$ mesons produced in the processes to be
$0^+$(even)$^{++}$, that is, $f_{J={\rm even}}$.
For higher energies, we invoke a quark model.
In the leading order calculations~\cite{bl,bc},
the $\pi^0\pi^0$ cross section is predicted to be 
much smaller: around 0.03-0.06 of $\pi^+\pi^-$.
However, higher-order or non-perturbative QCD effects can 
modify the ratio~\cite{handbag}. 
Analyses of energy and angular distributions of the cross sections
are also essential to test the validity of QCD models.
Also the $\chi_{cJ}$ charmonia are expected to contribute
to this process.

We previously presented the differential cross section, 
$d\sigma/d|\cos \theta^*|$,
for  $\gamma \gamma \to \pi^+ \pi^-$ in the low and 
high energy regions~\cite{mori, nkzw}, 
and here we report the results for $\gamma \gamma \to \pi^0 \pi^0$ for
a wide two-photon center-of-mass (c.m.)
energy range from 0.6 to 4.0~GeV, and a c.m. angular range, 
$|\cos \theta^*| <0.8$
based on the data from Belle with integrated luminosity of 
95~fb$^{-1}$~\cite{prdpaper,pi0pi0}.
 We  also make a preliminary discussion in the higher energy region
based on the results from 
%the higher-statistics data corresponding to 
the integrated luminosity of 223~fb$^{-1}$~\cite{conf813}, which is three orders
of magnitude more of previous measurements.

\section{Measurement of {\boldmath $\gamma \gamma \to \pi^0\pi^0$}~$^{[6]}$} 

We use data collected with the Belle detector~\cite{belle} 
at the KEKB asymmetric-energy $e^+e^-$ collider~\cite{kekb}. 
The energy of the accelerator was set at or near 10.58~GeV.
%for the greater part
%of the data and at several different energies in 10.30 - 10.87~GeV 
%for a small fraction of the data.
The analysis is made in the ``zero-tag'' mode, where
neither the recoil electron nor positron is detected. 
We restrict the virtuality of the incident photons to be small
by imposing a requirement of a strict transverse-momentum  balance 
along the beam axis of the final-state hadronic system.

We collect events with multiple photons without 
any reconstructed tracks triggered by activity in the 
electromagnetic calorimeter, where the total energy exceeding
1.15~GeV or four or more energy clusters with a threshold around 
110~MeV are required. 
Just two reconstructed neutral pions are required. 
We apply a  transverse-momentum ($p_t$) 
balance cut for the two-pion system in the $e^+e^-$ c.m. frame,
$|\sum \mbox{\boldmath$p$}_t^*| < 0.05$~GeV/$c$.

%%\subsection{Candidate events and background subtractions}
After the event selection, we find about $1.26 \times 10^6$  
candidates for $\gamma \gamma \to \pi^0 \pi^0$ from the 95~fb$^{-1}$ data.
We can separate background events using 
transverse-momentum distributions of the candidate events
with fitting them to the signal and background components.
The invariant-mass resolution is estimated using 
signal Monte Carlo events
and the experimental $p_t$-balance distributions. 
By modeling this smearing effect of the resolution,
we apply an unfolding procedure to obtain the corrected invariant-mass
distribution between 0.9 and 2.4~GeV for each angular bin. 

We derive $d\sigma/d|\cos \theta^*|$
in 16 angular bins in $|\cos \theta^*|<0.8$. The result after 
integration over the angle is shown in Fig.~1.
Data points in the energy region where the contribution from
$\chi_{cJ}$ charmonia is dominant are removed, 
because we cannot separate the components from
$\chi_{c0}$, $\chi_{c2}$  and the continuum in a 
model-independent way, due to the finite mass resolution and insufficient
statistics (they are discussed in detail in the next section). 

We estimate the systematic errors for
the integrated cross sections, considering various sources
including trigger efficiency, background subtraction,
$\pi^0$ reconstruction efficiency etc. 
The typical error size in the intermediate and high energy regions 
are 10 - 11\%.
The results in the energy region below 1.0~GeV have larger errors.

We find prominent resonant structures near 0.98~GeV,
1.27~GeV, 1.65~GeV and 1.95~GeV. The first two are
from $f_0(980)$ and $f_2(1270)$, respectively, which
are also observed in the $\gamma\gamma \to \pi^+\pi^-$ process~\cite{mori}.
The latter two cannot readily be assigned to any well-known
states.
 A general behavior of amplitudes is studied by fitting the differential 
cross sections
in a simple model, which includes the S, D$_0$ and D$_2$ waves that 
are parametrized as
smooth background and resonances (Fig.~2). 
We obtain a reasonable fit with the
$f_2(1270)$ and $f_0(980)$ consistent with previous measurements.
The S wave prefers to have 
another resonance-like contribution in the mass region, 1.2-1.4~GeV/$c^2$.

\begin{figure*}[t]
\begin{minipage}[t]{82mm}
\begin{center}
\includegraphics[width=0.68\columnwidth]{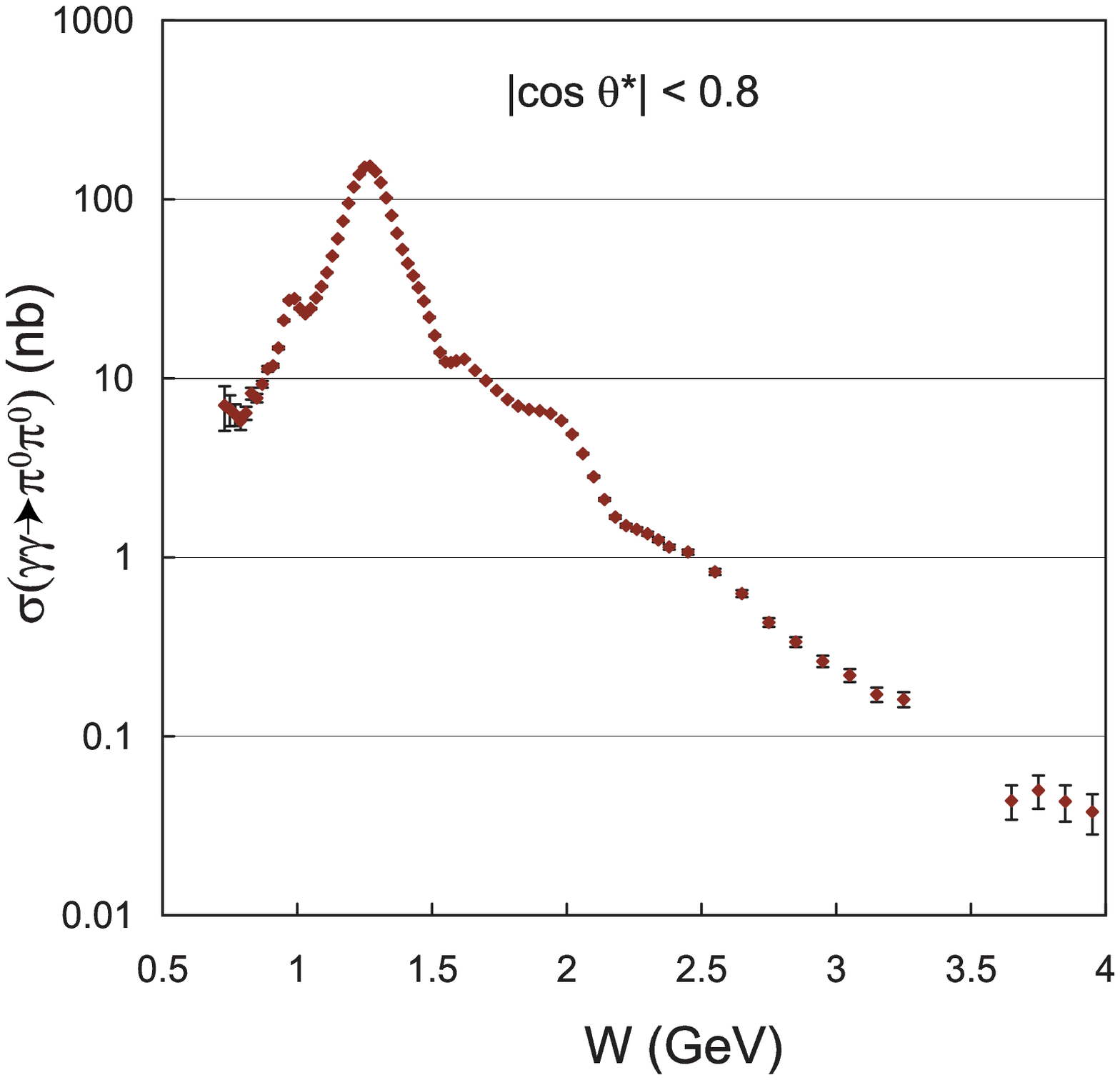}
\caption{The energy dependence of the cross section for 
$\gamma \gamma \to \pi^0\pi^0$ integrated over the angular 
range in $|\cos \theta^*|<0.8$.}
\vspace{2mm}
\end{center}
%\end{figure*}
\end{minipage}
\hspace{7mm}
\begin{minipage}[t]{82mm}
%\begin{figure*}[t]
\begin{center}
\includegraphics[width=0.68\columnwidth]{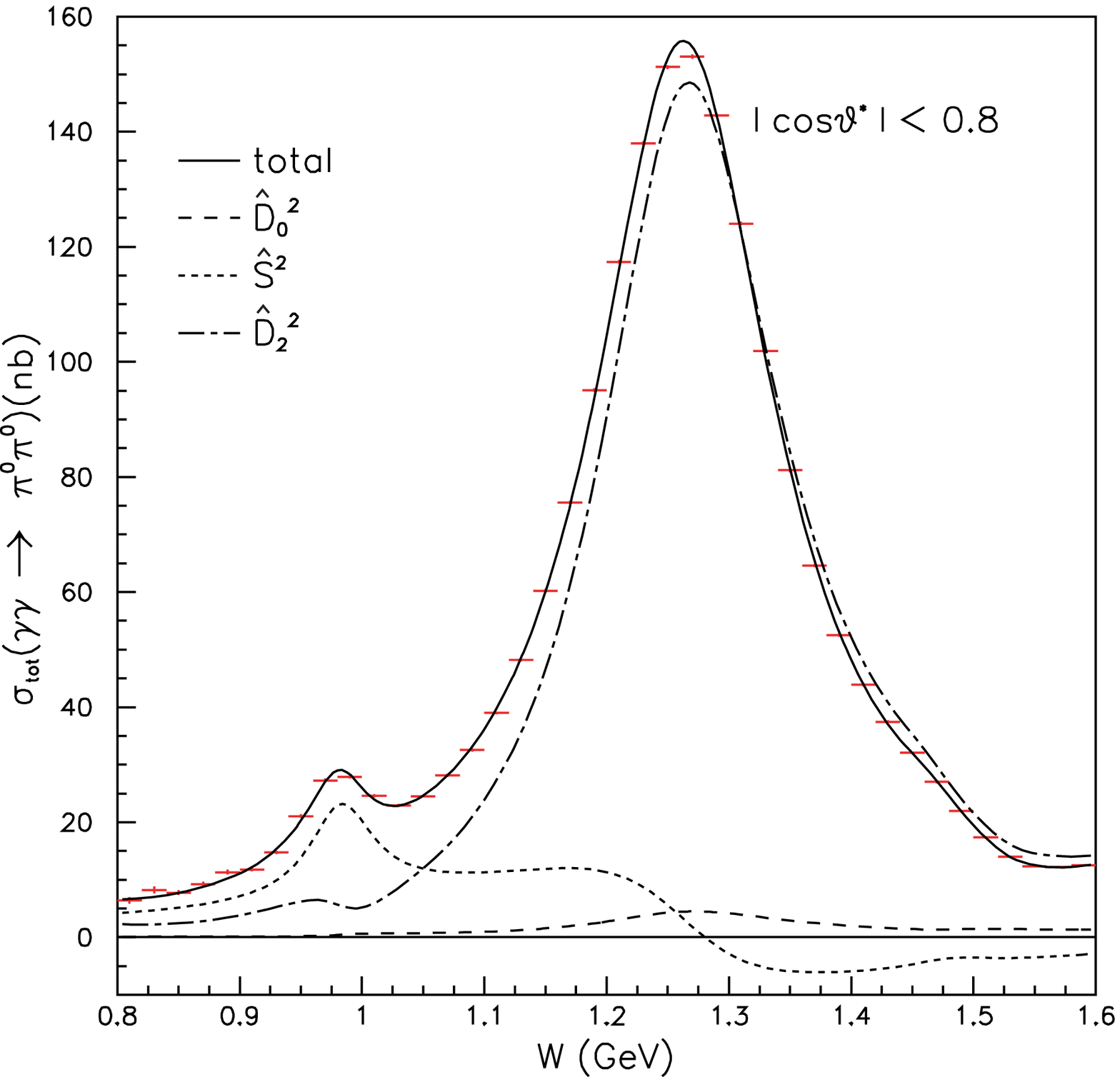}
\caption{Total cross section ( $|\cos \theta^*|<0.8$)
and results of parametrization. The contributions of partial
components are described in Ref.~\cite{prdpaper}.}
\end{center}
\end{minipage}
\end{figure*}

\section{Preliminary Analyses of the higher energy region and summary} 

We show the results from preliminary analyses based on the higher
statistical data sample corresponding to the integrated luminosity of
223~fb$^{-1}$~\cite{conf813}.
We find a signature of resonance-like structure near the $\chi_{cJ}$ 
charmonium region, 3.3-3.6~GeV. We fit the distribution of the yield
for the range 2.8~GeV$<W<4.0$~GeV
and $|\cos \theta^*|<0.4$ with a function including
the contributions from the $\chi_{c0}$ and $\chi_{c2}$
and a smooth contribution from the continuum, taking into account
partial interference between them.  The fit is shown in Fig.~3.
The mass and width of the charmonia are fixed to the world 
averages~\cite{pdg2008}. 
We obtain the products of the two-photon decay
width and the branching fraction,
$\Gamma_{\gamma\gamma}(\chi_{cJ}){\cal B}(\chi_{cJ} \to
\pi^0\pi^0) = 9.9^{+5.8}_{-4.0} \pm 1.0 $~eV and
$0.48 \pm 0.18 \pm 0.05 \pm 0.14$~eV, for $\chi_{c0}$ and $\chi_{c2}$,
respectively.
The first and second errors are for statistical and systematic, respectively,
and the last error for the $\chi_{c2}$ is from the uncertainty of 
the interference of the  $\chi_{c2}$ with the continuum component.  
These results are in agreement with the corresponding
world averages~\cite{pdg2008} and the charged pion-pair decays
from our measurement~\cite{nkzw}, taking isospin invariance into account. 

We show the cross section ratio between $\gamma \gamma \to \pi^0\pi^0$
and  $\gamma \gamma \to \pi^+\pi^-$~\cite{nkzw} for $|\cos \theta^*|<0.6$ and
2.4~GeV~$<W<4.1$~GeV, in Fig.~4(a). In the figure,
the error bars are statistical only, and each cross section
measurement has typically a 10\% systematic error.
The data points of  $\gamma \gamma \to \pi^+\pi^-$ above the charmonium 
masses have larger systematic error, $\sim$ 25\%. Subtractions of the
charmonium components are applied for these plots.

The cross-section ratio, neutral to charged, seems to have no big 
energy dependence for $W >2.7$~GeV and has an average ratio 
$\sigma(\pi^0\pi^0)/\sigma(\pi^+\pi^-)=
0.32 \pm 0.03 \pm 0.05$. 
This ratio is larger than the prediction from the lowest-order QCD
calculation~\cite{bl,bc}.
The angular distributions are also studied in the energy range above 1.4~GeV, 
and some implications to a {\it spin}-4 resonance and QCD models are discussed.

%\begin{multicols}{2}
\begin{figure*}[t]
\begin{minipage}[t]{82mm}
\begin{center}
\includegraphics[width=0.92\columnwidth]{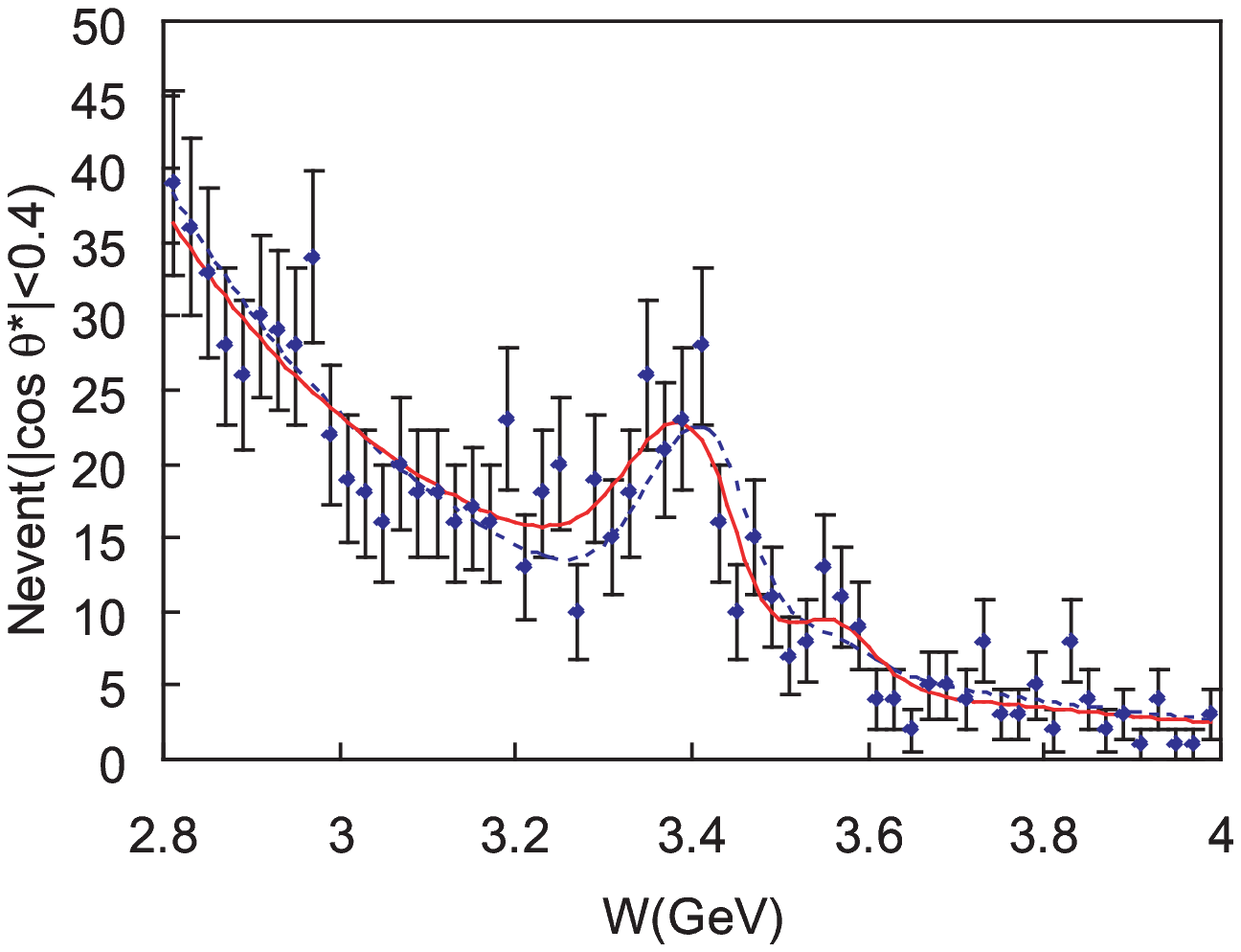}
\caption{Invariant-mass distribution of the $\pi^0\pi^0$ candidate 
events for $|\cos \theta^*|<0.4$ near the charmonium region. 
The solid (dashed) curves show the fits with (without) the interference 
of $\chi_{c0}$, respectively.}
\end{center}
%\end{figure*}
\end{minipage}
\hspace{7mm}
\begin{minipage}[t]{82mm}
%\begin{figure*}[t]
\begin{center}
\includegraphics[width=0.60\columnwidth]{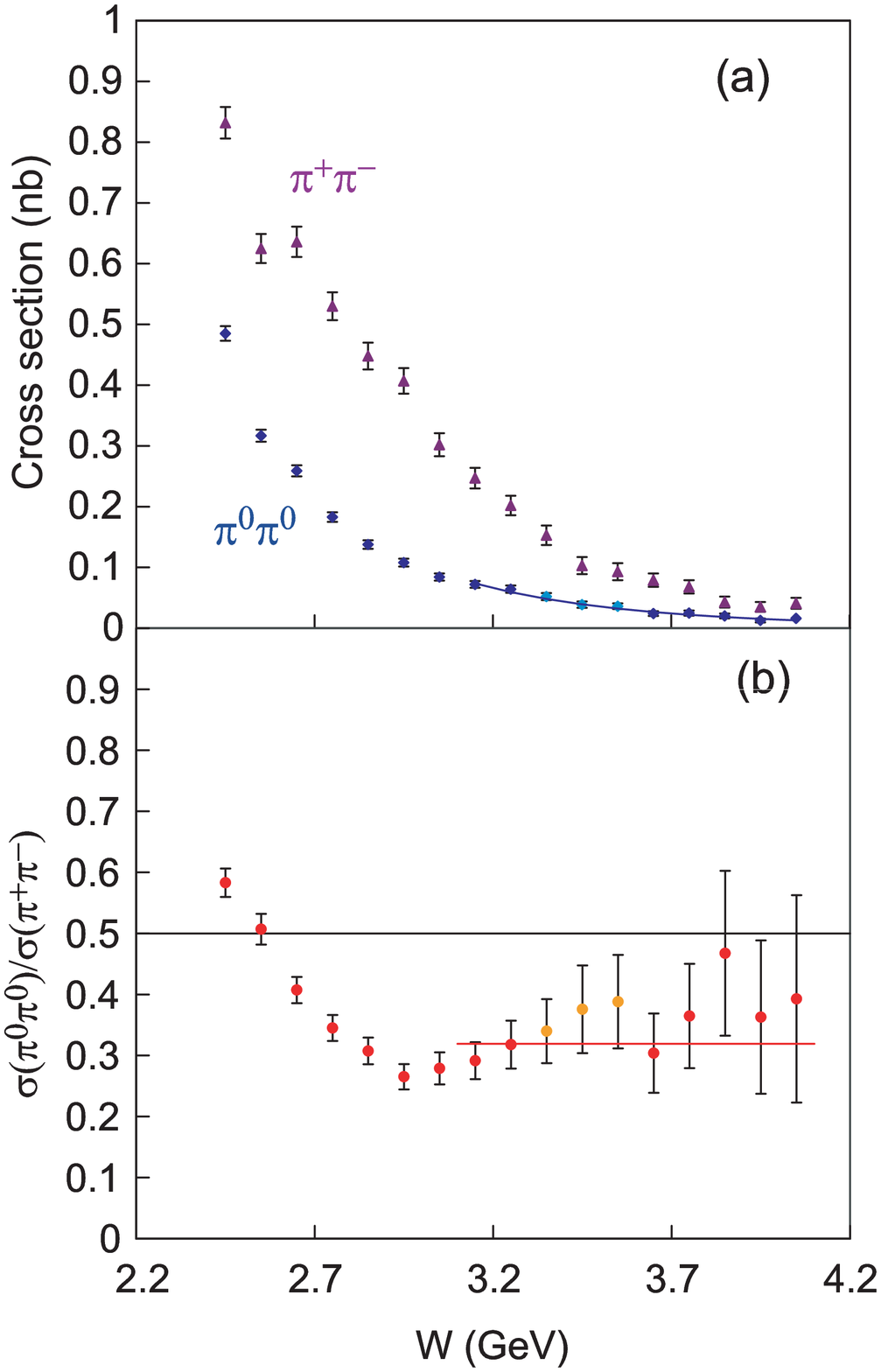}
\caption{(a) Cross sections for the $\gamma \gamma \to \pi^0 \pi^0$
and  $\gamma \gamma \to \pi^+ \pi^-$ processes for $|\cos \theta^*|<0.6$ 
and (b) their ratio. The lines are the fits to 
the result in the region indicated.
% (the data in the charmonium region are not used).
}
\end{center}
\end{minipage}
\end{figure*}
%\end{multicols}

%\section{Summary}
To summarize, we have measured the cross section of the process
$\gamma \gamma \to \pi^0\pi^0$ for the c.m. energy and angular regions,
0.60~GeV $< W < 4.1$~GeV and $|\cos \theta^*|<0.8$.
We observe some resonant structures, including the $f_0(980)$
and $\chi_{c0}$. 
%%We find that the angular dependence changes drastically at around
%%$W= 2.0$~GeV. 
The ratio of the cross sections of 
$\gamma \gamma \to \pi^0\pi^0$ and $\gamma \gamma \to \pi^+\pi^-$ 
in the 3 GeV region is obtained.

\clearpage

\end{document}